\documentstyle[preprint,prb,aps]{revtex}
\tighten
\begin{document}
\draft
\preprint{\today}

\title{Optical absorption in paired correlated random lattices}

\author{Francisco Dom\'{\i}nguez-Adame and Enrique Maci\'{a}$^*$}

\address{Departamento de F\'{\i}sica de Materiales,
Facultad de F\'{\i}sicas, Universidad Complutense,
E-28040 Madrid, Spain}

\author{Angel S\'anchez}

\address{Theoretical Division and Center for Nonlinear Studies,
Los Alamos National Laboratory,\\ Los Alamos, NM 87545\\ and\\
Escuela Polit\'ecnica Superior, Universidad Carlos III de Madrid, \\
C./ Butarque, 15, E-28911 Legan\'es, Madrid, Spain}

\maketitle

\begin{abstract}

Optical absorption in a random one-dimensional lattice in the presence
of paired correlated disorder is studied.  The absorption line shape
is evaluated by solving the microscopic equations of motion of the
Frenkel-exciton problem in the lattice.  We find that paired correlation
causes the occurrence of well-defined characteristic lines in the
absorption spectra which clearly differ from the case of unpaired
correlation.  The behavior of the absorption lines as a function of
defect concentration is studied in detail.  We also show how exciton
dynamics can be inferred from experimental data by deriving an
analytical expression relating the energy and intensity of lines to the
model parameters.

\end{abstract}


\narrowtext

\section{Introduction}

During the last few years, it is being realized that {\em correlated
disorder} has profound effects in quasiparticle dynamics and produces a
variety of complex and new phenomena.  The notion of correlated disorder
implies that certain physical parameters in random systems are not
completely independent within a given correlation length, thus leading
to a competition between short-range order effects and the underlying
long-range disorder.  It is well known, for instance, that delocalized
electrons \cite{Dunlap,Wu,Flores,PRBKP} and vibrations \cite{nozotro}
appear in one-dimensional random systems with correlated disorder.
Following the above line of research, we have recently carried out the
time-domain analysis of trapping of Frenkel excitons in one-dimensional
systems with paired correlated traps, randomly placed in an otherwise
perfect lattice. \cite{PRBF} Results were compared with the dynamics of
one-dimensional lattices whose excited states are also Frenkel excitons,
but in which transport is affected by the presence of unpaired traps.
\cite{Huber1} We found that pairing of traps leads to a slowdown of the
trapping rate due to the occurrence of larger segments of the lattice
which are free of traps.  In view of these results, the question arises
as to whether the fact that disorder is correlated will affect the
optical absorption spectrum of the lattice.  If so, it would be
interesting to find a relationship between the characteristic absorption
lines and the model parameters in order to compare with experimental
data.  In this paper we aim to answer these questions by numerical
simulation of optical spectra.  We believe that the interest of this
work is beyond the formal study of the exciton dynamics: Indeed, our
treatment may help describe the optical properties of some polymers with
active molecules, provided that interchain interactions become much
smaller that intrachain interactions so that a one-dimensional model can
be applied as a first approximation.

\section{Model}

The Hamiltonian for the Frenkel-exciton problem can be written in the
tight-binding form with nearest-neighbour interactions as follows (we
use units such that $\hbar=1$)
\begin{equation}
{\cal H}= \sum_{k}\> V_{k} a_{k}^{\dag}a_{k} +
T\sum_{k}\>(a_{k}^{\dag}a_{k+1}+a_{k+1}^{\dag}a_{k}).
\label{perfectH}
\end{equation}
Here $a_{k}$ and $a_{k}^{\dag}$ are the exciton annhilitation and
creation operators, respectively. $V_k$ is the transition frequency at
site $k$ and $T$ is the nearest-neighbor coupling, which is assumed to
be constant in the whole lattice.  The paired correlated disorder is
introduced as follows: Transition frequencies can only take on two
values, $V$ and $V'$, with the additional constraint that $V'$ appears
only in pairs of neigbouring sites (the correlation length is rouhgly
the lattice spacing).  Such a pair of sites will be called {\em dimer
defect}.  In addition, we define the defect concentration $c$ as the
ratio between the number of sites with transition frequency $V'$ and the
total number of sites in the lattice.

The line shape $I(E)$ of an optical-absorption process in which a single
exciton is created in a lattice with $N$ sites is given by
(see Ref.~\onlinecite{Huber2} for details)
\begin{equation}
I(E)=-\,{2\over \pi N} \int_0^\infty\> dt\, e^{-\alpha t} \sin (Et)
\,\mbox{\rm Im} \left( \sum_k\> G_k(t) \right),
\label{line}
\end{equation}
where the factor $\exp (-\alpha t)$ takes into account the broadening
due to the Lorentzian instrumental resolution function of width
$\alpha$.  The correlation functions $G_{k}(t)$ obey the equations of
motion
\begin{equation}
i{d\over dt} G_{k}(t) = \sum_{j}\> H_{kj} G_{j}(t),
\label{motion}
\end{equation}
with the initial conditions $G_{k}(0)=1$.  The diagonal elements of the
tridiagonal matrix $H_{jk}$ are $V_{k}$ whereas off-diagonal elements
are simply given by $T$.  Once these equations of motion are solved, the
line shape is found using Eq.~(\ref{line}).

\section{Results and discussions}

We have numerically solved the equation of motion (\ref{motion}) for
chains of $N=2\, 000$ sites using an implicit integration scheme.  In
order to minimize end effects, spatial periodic boundary conditions are
introduced.  Once the functions $G_k(t)$ are known, the line shape
$I(E)$ is evaluated by means of (\ref{line}).  Energy will be measured
in units of $T$ whereas time will be expressed in units of $T^{-1}$.
Since we are mainly interested in the effects due to the presence of
correlated disorder in the random system rather than in the effects of
the different parameters on the optical absorption process, we will fix
the values of $V$, $V'$ and $T$, focusing our attention on the defect
concentration $c$.  Thus, we have set $V=4$, $V'=10$ and $T=-1$
henceafter as representative values.  The width of the instrumental
resolution was $\alpha=0.5$.  The maximum integration time and the
integration time step were $16$ and $8\times 10^{-3}$, respectively;
larger maximum integration times or smaller time steps led to the same
general results.  The defect concentration $c$ ranged from $0.01$ up to
$0.9$ and for each lattice realization a random distribution of dimer
defects was chosen.  In addition, lattices with unpaired defects have
been studied and compared with lattices containing the same fraction of
paired defects.  This enables us to separate the effects merely due to
optical absorption in one-dimension from those which manifest the
peculiarities of the correlation between random defects.

In absence of defects the spectrum is a single Lorentzian line
centered at $E=V+2T$, which with our choice of parameters is $E=2$.
When defects are introduced in the lattice, a broadening of this main
line is observed accompanied by a shift of its position to higher
energies on increasing defect concentration.  Keeping these general
results in mind we now proceed to discuss the main features of the
spectra obtained when defects are present in the system.  The
obtained results are shown in Figs.~\ref{fig1}-\ref{fig3}, which
respectively correspond to an increasing defect concentration.  From a
close inspection of these figures several conclusions can be drawn.
First of all, we observe that, in addition to the main absorption line
located close to $E=2$, several satellites appear in the spectra.  These
satellites become more intense as the defect concentration increases.
Such lines are related to the presence of defects in the lattice, and
their positions only depend on the particular kind of disorder (paired
or unpaired).  On increasing defect concentration, the high-energy lines
increase at the expenses of the main line (compare
Figs.~\ref{fig1}-\ref{fig3}).  For low defect concentration (see
Fig.~\ref{fig1}), two lines appear at about $E=9.0$ and $E=10.2$, whose
positions are almost independent of $c$.  The line at $E=9.0$ is more
intense for lattices with dimer defects whereas the line at $E=10.2$ is
weaker in that case.  In addition, a small shoulder appers at about
$E=3.8$, and it becomes more intense in the case of unpaired lattices.
On increasing defect concentration, this shoulder is clearly resolved
and a new line develops at around $E=8.5$ (see Fig.~\ref{fig2}).  The
intensity of this new line rapidly increases with $c$ and shifts to the
low-energy region of the spectra (see Fig.~\ref{fig3}), until it is
placed close to $E=V'+2T$ ($E=8$ with the choosen parameters).

Firstly we discuss the shift of the main absorption line toward higher
energies on increasing defect concentration.  In the case of unpaired
defects this shift $\Delta E$ depends linearly on defect concentration
$c$; using a least squares fit we have found that $\Delta E \approx 0.92
c$ in the range $c=0.01$ up to $c=0.4$.  This behavior is in agreement
with the average-T-matrix approximation (ATA), which predicts that the
shift increases linearly with defect concentration in the absence of
correlations. \cite{Huber2} On the contrary, pairing of defects causes a
quadratic dependence on the defect concentration of the form $\Delta E
\approx 1.20 c^2$.  It is worth mentioning that the line shift $\Delta
E$ for the paired correlated lattice is always smaller than $\Delta E$
for the unpaired one, for the same defect concentration $c$ in the range
considered.  This is an important result since it suggests that pairing
of defects reduces the effects of disorder on the Frenkel-exciton
dynamics.  The physical explanation of this behavior is based on the
particular spatial distribution displayed by the defects when pairing is
present.  It becomes clear that the average length of segments which are
free of defetcs is larger when correlation is present since defects
always appear in pairs.  Therefore, pairing of defects implies a smaller
disruption of the lattice and consequently the random distribution of
such defects causes less influence on the exciton dynamics.

Now we focus our attention on the satellite lines present in the
spectra.  In order to study the results from a quantitative
point of view, we consider the following two center problem, which
describes the optical absorption spectrum of two isolated, coupled
defect sites
\begin{eqnarray}
i{d\over dt} G_{1}(t) &=& V_1G_{1}(t)+T G_{2}(t), \nonumber \\
i{d\over dt} G_{2}(t) &=& V_2G_{2}(t)+T G_{1}(t). \nonumber
\end{eqnarray}
These equations of motion can be easily solved using the initial
conditions $G_1(0)=G_2(0)=1$.  Inserting the result in (\ref{line}) and
taking $\alpha=0$ for the sake of simplicity, one obtains
\begin{equation}
I(E)= I_{+}\delta (E-E_{+})+ I_{-}\delta (E-E_{-}),
\label{spectratwo}
\end{equation}
where the intensity of each component is given by
\begin{equation}
I_{\pm}={1\over 2} \pm {1\over 2}\left[1+\left(
{V_1-V_2\over 2T} \right)^2\right]^{-1/2} \label{i} \nonumber
\end{equation}
and centered at
\begin{equation}
E_{\pm}={V_1+V_2\over 2}\pm T\sqrt{1+\left( {V_1-V_2\over 2T}\right)^2}.
\label{e}
\nonumber
\end{equation}
 From the Eqs.\,(\ref{i}) and (\ref{e}) we obtain the following possible
scenarios.  If the two transition frequencies of defects are
$V_1$=$V_2$=$V'$, the intensity $I_{-}$ vanishes due to the fact that
only the dipole matrix element involving symmetric states contribute to
the optical absorption spectrum.  In this case the spectrum exhibits a
single line centered at $E_{+}=V'+T=9.0$.  On the contrary, if the
transition frequencies are different (say $V_1=V$ and $V_2=V'$) the
optical spectrum presents two components centered at $E_{\pm}=(V+V')/2
\pm T\sqrt{1+(V-V')^2/4T^2}=7\mp \sqrt{10}$ so that $E_{-}\sim 10.2$ and
$E_{+}\sim 3.8$.

With this analytical results at hand, the quantitative interpretation of
the obtained spectra is straightforward since each absorption line can
be related to a specific kind of defect in the lattice.  We shall
consider, in first place, the case of low defect concentration
(Fig.~\ref{fig1}).  Besides the main line close to $E=2$, characteristic
satellites are clearly visible at about $E=10.2$ (more intense in
unpaired lattices) or $E=9.0$ (more intense in paired lattices).  These
values agree very well will our theoretical predictions shown above: The
line at $E=10.2$ is related to centers $V_1=V$ and $V_2=V'$, and the
number of such center is greater in unpaired lattices, whereas the line
at $E=9.0$ is related to dimer defects and the probability of finding
such defecs in unpaired lattices is actually very small at low $c$.
Analogous comments hold for the line centered at $E=3.8$ since it
originates from the same defects as the line centered at $E=10.2$.  On
further increasing defect concentration to intermediate ranges ($c\sim
0.5$) these characteristic lines continue to dominate the high energy
region of the absorption spectra (Fig.~\ref{fig2}).  Ultimately, in the
limit $c \rightarrow 1$, the absorption spectrum is dominated again by a
single line centered at $E=V'+2T=8$, as it corresponds for a lattice
enterely composed of defect sites.  This trend is clearly observed in
our spectra for both unpaired and paired lattices (Fig.~\ref{fig3}).

To seek further confirmation of the above explanations, we have
investigated the intensity dependence on the defect concentration for
the lines centered at $E=9.0$ and $E=10.2$.  Results are shown in
Fig.~\ref{fig4}.  The intensity of the line at $E=9.0$ increases almost
linearly at low defect concentration in the case of paired lattices, as
shown in Fig.~\ref{fig4}(a).  This is to be expected since such defects
are introduced on purpose in the lattice.  On the contrary, the
intensity of this line increases quadratically with defect concentration
in unpaired lattices since the probability of finding dimer defects in
such uncorrelated lattices also increases as $c^2$.  Concerning the line
centered at $E=10.2$, its intensity increases sublinearly with defect
concentration in the case of unpaired lattices, as shown in
Fig.~\ref{fig4}(b).  This is due to the fact that on increasing defect
concentration dimer defects appear and, consequently, the number of
centers $V_1=V$ and $V_2=V'$ cannot increase linearly.  In the case of
paired lattices the intensity of this line increases almost linearly
since the number of centers $V_1=V$ and $V_2=V'$ equals the number of
dimer defects whenever the probability of pairing of such defects
remains small at low defect concentration.  Therefore, we are led to
conclude that our interpretation of all relevant lines in the optical
spectra is correct, and those spectra show indeed clear differences when
coming from paired or unpaired disorder.

\section{Conclusions}

In summary, we have studied the absorption spectrum corresponding to the
Frenkel-exciton Hamiltonian for a random system in which a correlation
length of the order of the lattice spacing is intentionally introduced,
so that defects can only appear in pairs.  By comparing the obtained
spectra with that corresponding to the uncorrelated random system we are
able to identify characteristic absorption lines which can be directly
related to the presence of dimer defects.  Furthermore, we have obtained
analytical expressions which explain our synthetic spectra very well,
and relate microscopic system parameters, like transition frequencies or
defect concentration, to experimental data, like position of the lines.
Our results show that the presence of correlation effects in random
systems can be readily determined from analysis of the absorption
spectra of these samples, following the lines of reasoning presented in
this work.

\acknowledgments

The authors thank Bianchi M\'{e}ndez for a critical reading of the
manuscript.  A.S.\ is supported by a Ministerio de Educaci\'on y Ciencia
(Spain)/Fulbright postdoctoral scholarship, by Direcci\'on General de
Investigaci\'on Cient\'\i fica y T\'ecnica (Spain) through project
PB92-0378, and by the European Community ({\em Network}\/ on Nonlinear
Spatio-Temporal Structures in Semiconductor, Fluids, and Oscillator
Ensembles).  Work at Los Alamos is performed under the auspices of the
U.S.\ Deparment of Energy.


\begin{figure}
\caption{Absorption spectra for one dimensional random lattices
with concentration of paired defects $c=0.1$ (solid line) and $c=0.2$
(short-dashed line). Results are compared to lattices with the same
concentration of unpaired defects $c=0.1$ (long-dashed line) and $c=0.2$
(dot-dashed line). All spectra have the same area. Note the
magnification factor of the high-energy part of the spectra.}
\label{fig1}
\end{figure}

\begin{figure}
\caption{Absorption spectra for one dimensional random lattices
with concentration of paired defects $c=0.3$ (solid line) and $c=0.4$
(short-dashed line). Results are compared to lattices with the same
concentration of unpaired defects $c=0.3$ (long-dashed line) and $c=0.4$
(dot-dashed line).}
\label{fig2}
\end{figure}

\begin{figure}
\caption{Absorption spectra for one dimensional random lattices
with concentration of paired defects $c=0.7$ (solid line) and $c=0.9$
(short-dashed line). Results are compared to lattices with the same
concentration of unpaired defects $c=0.7$ (long-dashed line) and $c=0.9$
(dot-dashed line).}
\label{fig3}
\end{figure}

\begin{figure}
\caption{Intensity of the absorption lines centered at a) $E=9.0$ and b)
$E=10.2$ as a function of dimer defect concentration in random
one-dimensional lattices with paired defects (full circles) and unpaired
defects (full triangles). Solid lines indicate the best least squares
fit.}
\label{fig4}
\end{figure}

\end{document}